\documentclass[12pt,preprint]{aastex}
\def\msun{\mbox{M$_\odot$}}
\def\zsun{\mbox{Z$_\odot$}}
\def\I {{\'\i}}
\def\HII{\ion{H}{2}}

\shorttitle{C, N, and O Galactic gradients}
\shortauthors{Carigi et al.}

\received{}
\begin{document}

\title{Carbon, Nitrogen, and Oxygen Galactic Gradients: 
A Solution to the Carbon Enrichment Problem}

\author{Leticia Carigi}
\affil{Instituto de Astronom\'{\i}a, Universidad Nacional Aut\'onoma de M\'exico, 
Apdo. Postal 70-264, M\'exico 04510 D.F., Mexico}
\email{carigi@astroscu.unam.mx}

\author{Manuel Peimbert}
\affil {Instituto de Astronom\'{\i}a, Universidad Nacional Aut\'onoma de M\'exico,
Apdo. Postal 70-264, M\'exico 04510 D.F., Mexico} 
\email{peimbert@astroscu.unam.mx}

\author{C\'esar Esteban}
\affil{Instituto de Astrof{\'\i}sica de Canarias, E-38200, La Laguna, Tenerife, Spain}
\email{cel@ll.iac.es}

\author{and}

\author{Jorge Garc\'{\i}a-Rojas}
\affil{Instituto de Astrof{\'\i}sica de Canarias, E-38200, La Laguna, Tenerife, Spain}
\email{jogarcia@ll.iac.es}

\begin{abstract}

Eleven models of Galactic chemical evolution, differing in the carbon, nitrogen, and
oxygen yields
adopted, have been computed to reproduce the Galactic 
O/H values obtained from \HII \ regions. 
All the models fit the oxygen gradient, but
only two models fit also the carbon gradient, those based on carbon
yields that
increase with metallicity due to stellar winds in massive  stars (MS) and
decrease with metallicity due to stellar winds in low and
intermediate mass stars (LIMS). The successful models also fit the C/O versus O/H
evolution history of the solar vicinity obtained from stellar observations. 
We also compare the present day N/H gradient and the N/O versus O/H 
and the C/Fe, N/Fe, O/Fe versus Fe/H 
evolution histories of the solar vicinity predicted by our two best models
with those derived from \HII \ regions and from  stellar observations.
While our two best models fit the C/H and O/H gradients as well as the C/O versus O/H
history, only Model 1 fits well the N/H gradient and the N/O values
for metal poor stars but fails to fit the N/H values for metal rich
stars.
Therefore we conclude that our two best models solve the C enrichment problem,
but that further work needs to be done on the N enrichment problem.
By adding the C and O production since the Sun was formed predicted by Models 1 and 2
to the observed solar values we find an excellent agreement with the O/H and C/H values of
the solar vicinity derived from \HII \ regions O and C recombination lines.
Our results are based on an IMF steeper than Salpeter's, a Salpeter like IMF predicts
C/H, N/H, and O/H ratios higher than observed.
One of the most important results of this paper is that the fraction of carbon due
to MS and LIMS in the interstellar medium is strongly
dependent on time and on the galactocentric distance; at present  
about half of the carbon in the interstellar medium of the solar vicinity has been 
produced by MS and half by LIMS.

\end {abstract}


\section{Introduction}

Many chemical evolution models have been recently made to explain the chemical
composition of the solar vicinity 
(e.g.
Henry, Edmunds, \& K\"oppen 2000,
 Liang, Zhao \& Shi 2001, Chiappini, Matteucci \& Meynet 2003a,
Chiappini, Romano, \& Matteucci 2003b, Akerman et al. 2004).
In addition a few  models have been computed to 
explain the behavior of C/O as a function of the distance 
to the Galactic center
(e.g. Hou,  Prantzos, \& Boissier 2000,
Carigi 2003, Chiappini et al. 2003b, Gavil\'an, Buell, \& Moll\'a 2005). 
In this paper the word
gradient of an abundance ratio will be used to denote the galactocentric slope and  
the absolute value in the solar vicinity.

Most of the Galactic chemical evolution models predict a similar
history for C/O versus O/H at the solar vicinity, but make different predictions
for the behavior of C/O at different Galactocentric distances.
All authors agree that both massive stars (MS) and 
low and intermediate mass stars (LIMS) play a significant role in 
the C production of the solar vicinity, nevertheless some authors
find that most of the C is due to MS (e.g. Carigi 2000, 2003; Henry et al. 2000)
while other authors find that most of the C is due to LIMS 
(e.g. Chiappini et al. 2003b). The different predictions on:
the C/H value in the solar vicinity,
the Galactic C/O gradient, 
and the relative importance of MS and LIMS
in the C production
are mainly due to the stellar chemical
evolution models  obtained with different C yields.
To discriminate among the sets of yields
we need additional observational constraints to those used before.
The C/H  and O/H values derived from \HII \ regions 
 at different galactocentric distances by
Esteban et al. (2004a, hereinafter Paper I), provide us
with the additional constraints necessary to study this problem.
In this paper we present
eleven different chemical evolution models for the Galaxy,
based on combinations of eight different stellar yields,
to try to fit the observed C/O gradient.

It is difficult to study the C enrichment of the Galaxy because C is produced
by MS and LIMS and the evolution of both types of objects depends on:
stellar winds, the convection treatment, and the $^{12}$C($\alpha$,$\gamma$)$^{16}$O rate,
and an exact treatment of these three ingredients of the models is not yet available
(e.g. El Eid et al. 2004, Herwig \& Austin 2004).
Due to these reasons many different estimates of the C yields for MS and LIMS are 
available in the literature.
We have called ``the C enrichment problem"  the difficulty of estimating the proper
C yields for MS and LIMS.
In this paper we explore different solutions to the C problem studying only 
the effect of the stellar winds on the value of C yields.

It is even more difficult to study the N enrichment of the Galaxy for the following
reasons: it is produced by LIMS and MS, it can have a primary or secondary
origin, and in general its abundance is considerably smaller than that of
C and O. This last point implies that an uncertainty in the secondary production of N 
will affect considerably more the N abundances than the C and O abundances.
Consequently on theoretical grounds the predicted O and C values are more
robust than the N values. {From} observations of \HII \ regions the abundances 
of C and O are of higher quality than those of N for the following reasons:
for O we observe all the stages of ionization, for carbon we observe only the 
C$^{++}$ fraction and an ionization correction factor, ICF, is needed to correct for the
unseen ions, fortunately for C the ICF is small. On the other hand for N
we observe only the N${^+}$ fraction
which in general is not the most abundant ion, and consequently large ICFs are
needed to obtain the total N abundance. Moreover the C and O abundances
used in this paper are based on recombination lines that are almost independent
of the temperature structure of the \HII \ regions, while the N${^+}$ abundances
are based on collisionally excited lines that do depend on the temperature
structure of the nebulae. To reach agreement between the different N/H observational
data as well as between the observations and the
predicted N/H values will be called ``the nitrogen enrichment problem".

We will use also the C/Fe, N/Fe, O/Fe versus Fe/H enrichment histories 
that provide us with  additional checks to the models presented in this paper.

\section{Observational Constraints}

Our models will be compared with unevolved F and G stars and Galactic  \HII \ regions.
In this work the data used as observational constraints are the following:
i) the C/H and O/H abundances from Galactic 
\HII \ regions presented in Paper I
and the N/H abundances presented in Table 1 and Figure 1 (Esteban et al. 2004b, 
Garc\I a-Rojas et al. 2004, and in preparation),
ii) the H, C, N, O, and Fe abundances from main sequence stars in the solar vicinity 
obtained by
Santos, Israelian, \& Mayor (2000), Takeda et al. (2001),
Gonz\'alez et al. (2001), Sadakane et al. (2002),
Akerman et al. (2004),  and
Israelian  et al. (2004), and
iii) the H, C, N, O, and Fe solar abundances 
(Asplund, Grevesse, \& Sauval 2005).

In Paper I  new C/H and O/H gaseous values for  eight \HII \ regions 
between 6 and 11 kpc are presented,
adopting a Galactocentric distance for the Sun of 8 kpc
(see Fig. 2 of Paper I).
These C/H and O/H values have been increased by 0.10 dex and 0.08 dex,
respectively, due to the fraction
of C and O embedded in dust grains (Esteban et al. 1998).
Based on these data the C/H, O/H and C/O slopes of the gradients are
$-0.103$, $-0.044$, and $-0.058$   dex kpc$^{-1}$, respectively.

The values of Table 1 were obtained from the Very Large Telescope and the
Ultraviolet Echelle Spectrograph (Esteban et al. 2004, Garc\I a-Rojas et al.
2004, and in preparation). The N$^+$ abundances were obtained from
collisionally excited lines taking into account the temperature structure
by adopting $t^2$ values different from 0.00 (see Peimbert 1967 for the
definition of $t^2$), the increase in the N/H
values due to this effect amounts to about 0.21 dex. The total nitrogen abundances
were obtained based on the ICF values derived from the models by Mathis
\& Rosa (1991), note that these ICF values yield N abundances about 0.11 dex
higher than those derived from the usual formulation by Peimbert and Costero (1969)
where it is assumed that N$^+$/O$^+$ is equal to N/O.

In Figure 1 we present the N/H, N/O, and N/C values versus Galactocentric
distance based on the N/H values presented in Table 1 and the C/H and O/H values
presented in Paper I. The linear fits to the data, giving equal weight
to each value, are:

12 + log(N/H) =   (8.517 $\pm$0.156)$-$(0.085$\pm$0.020)$R_G$,  
     
     log(N/O) =  ($-0.606 \pm$0.113)$-$(0.042$\pm$0.015)$R_G$,
       
     log(N/C) =  ($-0.979 \pm$0.190)$+$(0.017$\pm$0.024)$R_G$. 

It is not possible to estimate the total Fe abundances based on the Fe
gaseous abundances derived from \HII \ regions data because most of the
Fe atoms are embedded in dust grains (Esteban et al. 1998). Therefore
we will use only the Fe values from stars of the solar vicinity to compare 
with our models.

Akerman et al. (2004) present C/H,  O/H,  and Fe/H stellar values from 34 F
 and G dwarf stars of the Galactic halo
and combine their values with similar data from 19 disk stars.
Based on these data the C/O value in the solar vicinity
drops from 12+log(O/H) $\sim$ 6  to $\sim 7.7$ 
and then increases from 12+log(O/H) $\sim $ 7.7 to 8.8.

Israelian et al. (2004) present N/H, O/H, and Fe/H stellar values from
31 metal-poor stars and 15 metal rich stars.
Based on these data it is found that the N/O values in the solar vicinity
increase with O/H with the exception of  the N-rich star G64-12.
We did not consider those metal-poor stars with only an upper limit in N/H.
Santos et al. (2000), Takeda et al. (2001),
Gonz\'alez et al. (2001), and Sadakane et al. (2002)
present C/H stellar values for the metal rich stars by Israelian et al..

\section{Chemical Evolution Models}

All models are built to reproduce
the observed gas fraction distribution of the Galaxy
and the observed O/H \HII \ region values from 6 to 11 kpc
at 13 Gyr, the age of the model, the time elapsed since the
beginning of the formation of the Galaxy. The observed O/H values
might be extrapolated to the 4 to 16 kpc range. 
The models do not reach the central regions of the Galaxy because
the evolution of the central regions might not
correspond to an extrapolation
of the disk values to the center for the following reasons:
a) the bulge has a different chemical evolution history;
b) the effect of the bar has to be considered;
c) the extrapolation of the models to higher metallicities might not be 
correct due to saturation effects.

The chemical evolution models for the solar vicinity of Akerman et al. (2004) 
have been extended to 
follow the chemical history of the Galactic disk 
in the 4 kpc to 16 kpc range, under the following assumptions:

i) 
The total surface mass density adopted
as a function of time and Galactocentric distance $r$ is given by
\begin{equation}
\frac{d \sigma_{\rm gas+stars}}{dt} =
A(r) e^{-t/\tau_{\rm halo}} + B(r) e^{-(t-1 Gyr)/\tau_{\rm disk}},
\end{equation}
where the formation timescales
$\tau_{\rm halo}=0.5$ Gyr and $\tau_{\rm disk}=6 + (r/r_\odot - 1)8$ Gyr, 
 and the constants
$A(r)$ and $B(r)$ are chosen to match the present-day radial distribution 
of gas surface mass density (Prantzos 2003);
the total surface mass density is given by 
($\sigma_{tot}(r_\odot,t_g)=50 e^{-(r-8{\rm kpc})/3.5{\rm kpc}}$ \msun $pc^{-2}$),
where we have adopted the halo to disk mass ratio of the solar vicinity
for all Galactocentric distances.

ii)
The star formation rate is proportional to a power of $\sigma_{gas}$ and
   $\sigma_{tot}$: $ SFR(r,t) = \nu \ \sigma^{1.4}_{gas}(r,t) \ \sigma^{0.4}_{tot}(r,t)$, 
where
   $\nu$ is a constant in time and space.
In order to improve the agreement of the halo
to disk abrupt change in C/O at 12+log(O/H) $\sim$ 8, 
we assume a $\nu$ value five times higher
during the halo formation than that adopted for the disk.

A $\nu$ value for the disk of 0.016 
has been adopted  for all models, with 
the exception of those  models that assume the yields by Woosley \& Weaver (1995).
For these yields the adopted $\nu$ value is 0.019.

iii) 
The Initial Mass Function (IMF) adopted is  the one 
proposed by Kroupa, Tout \& Gilmore (1993, KTG)
in the mass interval given by  
0.01 $< m/\msun <$ 80.
The KTG IMF is a three power-law approximation, given by
IMF $\propto m^{-\alpha}$ with
$\alpha=-1.3$ for 0.01 - 0.5 \msun, 
$\alpha=-2.2$ for 0.5 - 1.0 \msun,
and
$\alpha=-2.7$ for 0.5 - 80 \msun.

iv) 
In order to study the contribution to the C, N, and O enrichment of the 
interstellar medium, ISM, due to stellar evolution we have assumed 
different sets of stellar yields,
all dependent on metallicity.

For massive stars (MS), those with $8 < m/\msun < 80$, we have used
the following five sets of yields:
a) Chieffi \& Limongi (2002) for initial metallicities (by mass) $Z = 0.0$;
b) Meynet \& Maeder (2002, MM02)
for  $Z = 10^{-5}$,
$Z = 0.004$, and $Z = 0.02$;
c) For all elements MM02 for $Z = 1 \times 10^{-5}$ and $Z = 0.004$,
for C and O Maeder (1992, M92) for $Z = 0.02$ (high mass-loss rate)
for N MM02 for $Z = 0.02$;
d) Woosley \& Weaver (1995, WW95) from $Z = 10^{-4}$ to $Z = 0.02$;
e) Portinari, Chiosi, \& Bressan (1998, PCB98) from $Z =0.0004$ to $Z = 0.05$.

For  low and intermediate mass stars (LIMS), those with $0.8 \leq m/\msun \leq 8$, 
we have used the following three sets of yields:
a) Meynet \& Maeder (2002, MM02) for $Z=1 \times 10^{-5}$; 
b) van den Hoek \& Groenewegen (1997, vdHG) from $Z=0.001$ to $Z=0.04$
with constant or variable $\eta$ 
(parameter that represents the importance of mass loss during the AGB phase)
as a function of $Z$, where in the first case $\eta=4$ for all $Z$ and
in the second case
$\eta=1$ for $Z = 0.001$, $\eta=2 $ for $Z = 0.004$, and
$\eta=4$ for all other  $Z$ values; and
c) Marigo, Bressan, \& Chiosi (1996, 1998), and Portinari et al.  (1998)
from $Z=0.004$ to $Z=0.02$, these sets of yields have been labelled MBCP.

For massive stars  we have used Fe yields by Woosley \& Weaver (1995)
(Models B, for 12 to 30 \msun; Models C, for 35 to 40 \msun).
For $m > 40 $ \msun, we have extrapolated the Fe yields for $m = 40$ \msun.

We have assumed that
5 \% of the stars with initial masses between 3 and 16 \msun \ are  binary systems which
explode as Type Ia SNe with the yields computed  by Thielemann, Nomoto \& Hashimoto (1993).
This fraction is needed to explain the metallicity distribution of local G and K dwarf stars.

For each set of yields, linear interpolations for different stellar masses
and metallicities were made.
For metallicities higher or lower than those available we adopted the yields predicted by
the highest or lowest $Z$ available, respectively.

The models fit  many observational constraints related to the chemical abundances,
for example: the total surface density, the infall rate, and the star formation rate
of the solar vicinity and the Galactic disk.

\section{Results}

\subsection{Carbon and oxygen}

The essence of this work is to explore the behavior of the C/O gradient in the Milky Way
and the C/O history of the solar vicinity.
It is important to note that
the O/H values depend on the O yields, the initial mass function, and the Galaxy
formation history and are well adjusted by all models.
While the C/O ratio depends mainly on the C yields, therefore permitting us
to discriminate among the C yields available.

In  Table 2, we present the following predictions of the models for the present time:
i) the C/O value at $r=r_\odot$, and
ii) the C/O value of the slope of the gradient for the 6 to 11 kpc
range, zone that corresponds to the observations of Paper I.
{From} this table, it can be noted that:

i) The C/O  value at the solar Galactocentric radius is reproduced by Models 1, 2, 9, 
and 10 that adopt  two kinds of stellar yields.
In Models 1, 2, and 9 we adopted the C and O yields for 
MS with high stellar winds (M92, PCB98)  and the C yields for LIMS that decrease
with $Z$ (MBCP or vdHG.var).
In Model 10 we adopted the C and O yields for
MS without stellar winds (WW95) and the C yields for LIMS that decrease
with $Z$ (vdHG.var).

ii) Models 3 to 8 and 11 predict for $r=r_\odot$ C/O values
0.1 to 0.4 dex lower than  observed.

iii) The C/O gradient is only reproduced by Models 1 to 4 that assume for massive stars
the C MM02 yields for $10^{-5} \le Z \le 0.004$ 
and the C M92 yields for $Z = 0.02 $.

iv) Models 5 to 11 predict C/O gradients flatter than observed.
For Models 5 to 8  the MS MM02 yields
with $Z = 0.02$  have winds with lower mass loss rates 
than those by M92, producing a smaller amount of C.
For Model 9 the C/O gradient is almost flat because the MS PCB98 yields
with $Z > 0.02$ include intense winds, which occur before
these stars synthesize C; therefore, their C  yields become similar
to those without stellar winds (Carigi, 2000).
For Models 10 and 11 the MS  WW95 yields do not include stellar winds and consequently
the C yields do not depend on $Z$.

v) The low C/O values predicted by Models 4 and 8 are mainly due to the MM02 yields for LIMS,
the lowest ones considered in this paper,
because  the MM02 yields did not extend to the thermal-pulse AGB phase,
consequently the third dredge-up and the hot-bottom burning stages were not included.
Moreover, the computations based on the MM02 yields for LIMS
are somewhat uncertain because the only yields available
are from 2 \msun \ to 7 \msun \ for $Z=10^{-5}$,
and for 3 \msun \ in for $Z=0.004$ and $Z=0.02$.

vi) Models 2 and 6, with vdHG.const yields, predict higher C/O values
than Models 3 and 7, with vdHG.var yields,
because the vdHG.const yields are computed with $\eta=4$ 
while  vdHG.var yields are computed with  $\eta$  between 1 and 4.
With higher $\eta$ values the stars lose more gas and the  AGB lifetimes become shorter,
reducing the C yields.

vii) The highest C/O values with the same MS yields were predicted by Models 1 and 5,
because the yields by MBCP for LIMS were computed with $\eta$ values lower than
those assumed by vdHG.

As mentioned in the Introduction the C enrichment is complex and depends on
many variables. Therefore, we decided to
evaluate the relative importance of MS and LIMS in the production of C
and to compute the fraction of C in the ISM due to both
types of stars.
In Table 3, we present the C processed and ejected to the ISM by MS, LIMS
and Type Ia SNe during the whole evolution of the solar vicinity.
{From} this table it can be noted that:
a) Models 1 and 2,
that are the two models that best fit the data, predict that MS produce between 48 \% and 50 \%
and LIMS produce between 50 \% and 47 \%  of the ISM  present-day carbon abundance,
b) the unsuccessful models predict that the C produced by MS varies from 36 \% to 75 \%
and the  C produced by LIMS varies from 62 \% to 26 \%,
c) the rest of the C is produced by type Ia SNe.
The relative importance of the C production of MS and LIMS
changes with time (Akerman et al. 2004) and
with Galactocentric radius.

In Figure 2 we present the cumulative C enrichment of the ISM of the solar vicinity
as a function of time and O/H for Models 1 and 2.
As can be seen from the figure MS dominate the C enrichment at early times,
12 + log(O/H) $< 8$. For latter times, 12 + log(O/H) $> 8$,
the contributions by  MS and LIMS become comparable.
At 12 + log(O/H) $\sim 8$ the contribution of LIMS to the C enrichment
increases because the LIMS formed in the halo end their evolution ejecting 
freshly made C, note that the low $Z$ LIMS eject more C than the high $Z$ ones.
For  12 + log(O/H) $> 8.5$  the relative decrease in C production by LIMS
is due to the higher contribution of the high $Z$ MS to the C production,
and the lower contribution of the high $Z$ LIMS.

In Table 4 we present the fraction of C due to MS and LIMS
in the ISM for Models 1 and 2 at  different Galactocentric distances.
As expected, the fraction of C present in the ISM produced by MS
increases with decreasing $r$ (for higher O/H values), because for MS the C yields
increase with $Z$, while for LIMS they decrease with $Z$.

To make a detailed comparison with the observed abundances we present in Table 5 the
C/H and O/H predicted values by Models 1 and 2 for the ISM at the time the Sun was formed
(4.57 Gyr ago) and at the present time.

In the left panel of Figure 3 we show the  C/O versus O/H enrichment
history for the solar vicinity predicted by Models 1 and 2. Also in this
figure we present three types of observations:
a) stellar values that are well fitted  by Models 1 and 2 
with the exception of those objects C/O $< - 0.8$ dex,
b) values for the two closest \HII \ regions in Galactocentric
distance to the solar vicinity, Orion and NGC 3576, that are perfectly fitted
by Model 1 and 2,
c) the solar values that are also perfectly fitted by Models 1 and 2.

In the right panels of Figure 3 we show the fit of Models 1 and 2
to the data presented in Paper I. All the models were constructed to reproduce
the O/H values derived from \HII \ regions, but Models 1 and 2 also produce 
an excellent fit to the C/H \HII \ region values
and a good fit to the C/O gradient, while the other models do not.
Note that for Galactocentric distances smaller than 6 kpc the C/O values
predicted by the models
start to saturate. 
Additional observational data as well as a model that includes
the behavior of the bulge are needed to study the regions with $r <$ 4 kpc.
For  Galactocentric distances, larger than 11 kpc, the C/O ratio predicted 
by Models 1 and 2 flattens
and again additional observations are needed to test the models.
The fit to the observations corresponds to the $6 < r({\rm kpc}) < 11$ range,
therefore Models 1 and 2 need to be tested for $r < 6$ kpc and $r > 11$ kpc
based on future observations.

\subsection{Nitrogen}

To constrain further the models of chemical evolution of the Galaxy and to
try to discriminate between Models 1 and 2 we decided to study the enrichment
of nitrogen during the history of the Galaxy.

In Models 1 and 2,
for $Z$ = \zsun, we have assumed N yields from MM02, computed with rotation and 
stellar winds; while for C and O we have used yields from M92, computed with 
stellar winds with a higher mass loss rate than those by MM02. 
The effects of stellar winds are 
very important for the C and O yields, but not for the N yields; while the 
effects due to rotation are more important for the N yields than for the C 
and O yields.

In Figure 4 we present the cumulative N enrichment of the ISM of the solar vicinity
as a function of time and O/H for Models 1 and 2.
As can be seen from the figure MS dominate the N enrichment at very early times,
12 + log(O/H) $< 6.5$. For latter times, 12 + log(O/H) $> 6.5$,
the contributions LIMS become more important.
The percentage of N by MS increases with
12 + log(O/H) higher than 8.1 due to the increasing importance of secondary production
in MS with increasing O/H.

In Table 6 we present the fraction of N due to MS and LIMS
in the ISM for Models 1 and 2 at  different Galactocentric distances.
As expected, the fraction of N present in the ISM produced by MS
increases with decreasing $r$ (for higher O/H values),
but in the computed range never produce a higher fraction of N
than LIMS.
The main reason for this increased production of N by MS
is that most of the N is produced by a secondary process.

In the left panel of Figure 5 we present the N/O versus O/H enrichment
history for the solar vicinity predicted by Models 1 and 2. Also in this
figure we present four types of observations: a) values for metal poor stars
that are well fitted by Model 1 with the exception of G6412 that appears to be
nitrogen rich, b) values for the two closest \HII \ regions in Galactocentric
distance to the solar vicinity, Orion and NGC 3576, that are perfectly fitted
by Model 1, c) the solar values that are intermediate between Models 1 and 2,
and that show an N/O excess relative to \HII \ regions of about 0.1 dex, and d)
values for metal rich stars that appear to be better adjusted by Model 2, but
that on average are about 0.3 dex higher than the N/O values determined from 
Orion and NGC 3576.
Also in Table 5  we present the
N/H values for ISM predicted  by Models 1 and 2  at the time the Sun was formed
 and at the present time.

In the right panels of Figure 5 we show the present behavior 
of N/H and N/O for the disk of the Galaxy predicted by Models 1 and 2.
In both cases the N/H and N/O \HII \ region values are in excellent agreement with 
Model 1 and in average about 0.2 dex smaller than those predicted by
Model 2.

In the left panel of Figure 6 we present the N/C versus O/H enrichment
history for the solar vicinity predicted by Models 1 and 2. Also in this
figure we present four types of observations: a) there are only four metal poor
stars in common between the data by Akerman et al. (2004), that present
C/H and O/H values, and that by
Israelian et al. (2004), that present N/H and O/H values, in Figure 6 
out of the four stars we present only two of them: HD 140283 and HD 194598, we
have an N/C value and two O/H values for each and therefore are plotted
twice, the other two stars are not plotted because one of them only has
an upper limit for the N/H value and the other is N rich and falls outside
the figure,
the observations fall between Models 1 and 2 but are closer to Model 2, 
b) values for  Orion and NGC 3576 that are perfectly fitted
by Model 1, c) the solar values that are intermediate between Models 1 and 2,
and that show an N/C excess relative to \HII \ regions of about 0.2 dex, and d)
values for metal rich stars that appear to be better adjusted by Model 2, but
that show N/C values from 0.1 to 0.6 dex higher than the N/C values 
determined for Orion and NGC 3576.

In the right panels of Figure 6 we show the present behavior 
of N/C for the disk of the Galaxy predicted by Models 1 and 2.
The \HII \ region values are in excellent agreement with 
Model 1 and are about 0.4 dex smaller than those predicted by
Model 2.

\subsection{Iron}

The history of Fe is different to that of O because in addition to
its production by SN of Type II, there is an important contribution due
to supernovae of Type Ia. Therefore Fe provides us not only with a consistency
check on the models but also with a direct constraint on the
production of Type Ia SNe and the Fe yields by massive stars.
Moreover C/Fe and N/Fe have been determined in many objects
and depend on  different physical parameters therefore  providing us with
additional information.
In Table 4  we present the
Fe/H values for ISM predicted  by Models 1 and 2  at the time the Sun was formed
 and at the present time.

In Figure 7 we present the C/Fe, N/Fe, and O/Fe observed data for
dwarf unevolved stars of the solar vicinity. The axis have been
normalized to the solar abundances by Asplund et al. (2005). 
In the three
panels of Figure 7 it is seen that metal rich stars, those in the
$-0.4$ to +0.4 [Fe/H] range, the C/Fe, N/Fe and O/Fe ratios fall above the
solar value.

Also in Figure 7 we present the C/Fe, N/Fe, and O/Fe versus Fe/H enrichment history for
the solar vicinity predicted by Models 1 and 2. The C/Fe and O/Fe versus Fe/H
histories for both models are practically the same and fit the solar values
reasonably well; for those stars with
Fe/H higher than $-0.6$ dex we find that their C/Fe and O/Fe values are about 
0.3 dex above the predictions by the model.

As can be seen in Figure 7 Model 1 predicts a N/Fe ratio about 0.15 dex
smaller than the solar value, while Model 2 predicts a N/Fe ratio about
0.2 dex higher than the solar value. Most of the other N/Fe star values presented
in Figure 7 lie between both models.

\section{Discussion}

\subsection{Observational constraints}

The O/H observations from \HII \  regions that have been used to fit the models are
those of Paper I that are based on recombination lines that to a very good
approximation are independent of the temperature structure of the nebulae. 
There are other determinations that are popular, and produce different results,
that are based on observations of
collisionally excited lines and that do depend strongly on the temperature
structure of the nebulae, we will mention three of them, those by 
Shaver et al. (1983), Deharveng et al. (2000), and Pilyugin, Ferrini, 
\& Shkvarun (2003).

The 12 + log O/H = 8.71 value for the solar vicinity by Shaver et al. (1983)
is similar to 8.77, our value; on the other hand they obtain a gradient for the O/H ratio
of $-0.07$ dex kpc$^{-1}$ for a solar Galactocentric distance of 10 kpc.
For a solar Galactocentric distance of 8.0 kpc, the one adopted by us, their gradient becomes  
$-0.086$ dex kpc$^{-1}$, considerably steeper than $-0.044$ dex kpc$^{-1}$, our value.
Our gradient implies lower O/H values for the inner regions of our galaxy
which is  consistent with recent suggestions that indicate that
the O abundances of \ion{H}{2} regions in the inner zones of external 
spiral galaxies are lower than previously
thought (Bresolin, Garnett, \& Kennicutt 2004). 

The slope of the O/H gradient presented in Paper I is in excellent agreement 
with that by Deharveng et al. (2000) that
amounts to $- 0.0395 \pm 0.0049$ dex kpc$^{-1}$, but
our O/H ratio for the ISM of the solar vicinity amounts to 12 + log O/H = 8.77, 
a value 0.29 dex higher than the 8.48 value
derived by them. The lower O/H value
by Deharveng et al. is due to their adoption of $t^2$ = 0.00 that reduces the
O/H value by 0.21 dex relative to the recombination line abundances and to the 
fraction of O trapped by dust grains that produces a further reduction of 0.08 dex.
The similar slope between both groups is probably due to the similar $t^2$
values  for Galactic \HII \ 
regions at different galactocentric distances that amount to about
0.035 (see Paper I and Table 2).

A similar situation to that by Deharveng et al. (2000) prevails with the O/H gradient 
by Pilyugin et al. (2003) that 
present a compilation of 71 observations of 13 galactic
 \HII \ regions by many authors and derive 
a slope for the 
gradient of $-0.048$ dex kpc$^{-1}$ and a 12 + log O/H = 8.50 value for the
solar vicinity. The slope of the gradient is in good agreement with that 
by Deharveng et al. and that presented in Paper I. But their O/H value
is 0.27 dex smaller than that obtained by us, again due to their adoption
of $t^2$ = 0.00 and the fraction of O trapped by dust grains not included by them.

The N/H \HII \  region values used by us are typically 0.3 dex higher than those
used by other authors.
The differences with most authors are due to the adoption by us
of $t^2 \not= 0.0 $ that yield on average an increase  of about
0.2 dex in the N/H values relative to those derived under the assumption that
$t^2$ = 0.00; an additional increase of about 0.1 dex comes from the adoption
of the ICFs by Mathis \& Rosa (1991) that in general are about 0.1 dex higher
than those derived under the assumption that N$^+$/O$^+$ = N/O.
Even higher N/H values for the \HII \  regions are needed to reach agreement with
the stellar values. We consider unlikely that a systematic difference in the adopted
$t^2$ and ICF values are responsible for the discrepancies between \HII \ regions
and stellar values.

The discussion of the N enrichment based on observations alone has an initial
problem that has not been solved, namely that \HII \ regions yield  N/O and N/C
values that in general are considerably smaller than those of metal rich 
unevolved dwarfs of the solar vicinity; the average difference amounts 
to 0.2 dex for N/O and to 0.3 dex for N/C. Furthermore the dispersions 
among the stellar data amounts to 0.6 dex for N/O and 0.5 dex for N/C 
while for the \HII \ regions the dispersions amount to 0.2 dex in N/O 
and 0.3 dex in N/C (see Figures 5 and 6). Part of the dispersion of the N/O values from 
\HII \ regions is due to the presence of an N/O gradient, on the other 
hand most of the dispersion in the N/O and N/C stellar data is not 
expected to be due to the dispersion of the ISM abundances at the time these stars were 
formed because the spreads in age and Galactocentric distances
are relative small. The N/O and N/C differences between the metal rich stars and 
the \HII \ regions need to be sorted out before an agreement between 
the chemical enrichment history of the Galaxy and the observations 
can be claimed.

A possible solution to the N differences between metal rich stars and \HII \ regions
could be due to a conversion of a small fraction of C into N after
the stars were formed. This effect, if present, would  help to explain not only
the large observational dispersion observed but also the larger N/H values
in these stars than in \HII \  regions. The fraction of C transformed into N 
needs to be small enough to avoid a substantial decrease in the C/H ratio,
since this ratio is well explained by the Galactic chemical evolution models.

There are two different sets of solar abundances:
a) the one by Asplund et al. (2005), based on a hydrodynamical 3D photospheric
model, given by:
H = 12.00 dex, C = 8.39 dex, N = 7.78 dex, O = 8.66 dex, and Fe = 7.45 dex, and
b) the one by Grevesse \& Sauval (1998), based on the classical 1D photospheric
model, given by: H = 12.00 dex, C = 8.52 dex, N = 7.92 dex, O = 8.83 dex,
and Fe = 7.50 dex. The standard solar models by Bahcall et al. (2004) 
are in good agreement with the
helioseismologically determined sound speed and density as a function
of solar radius, the depth of the convective zone, and the surface helium
abundance, as long as the models use the Grevesse \& Sauval solar abundances. The reasons
why the solar abundances by Grevesse \& Sauval produce a better fit than those by
Asplund et al.
to the standard solar model have to be found. To compare with our Galactic 
chemical evolution models we have used the Asplund et al. abundances.

The observational constraints from stellar data are not homogeneous.
The abundances for all stars but the Sun were determined assuming  
1D photospheric models.
Moreover some abundances have been computed adopting LTE models
and others adopting non-LTE models:
a) the O/H values from Akerman et al. (2004) were obtained in non LTE,
and the C/O values in LTE;
b) the O/H and N/H values from Israelian et al. (2004) were obtained in LTE;
and c) the solar abundances from Asplund et al. (2005) were obtained in non-LTE.

The Akerman et al. (2004)
 values show typical non LTE corrections of $-0.1$ to $-0.2$ dex for O/H,
but they cannot estimate those corrections for the C/O values.
Contrary to the opinion of Akerman et al., Israelian et al. (2004) argue that
the O/H and C/H values are almost independent of the non-LTE effects.
For metal poor stars the O/Fe values from Israelian et al. are higher than those
from Akerman et al.; the differences increase when Fe/H decreases and reach  values
of about 0.4 dex (see Figure 7).

From the solar experience it is expected that
3D models of metal rich stars will lower the O/H, C/H and N/H values but probably
not the N/C and N/O ratios.
If the N/C and N/O ratios are not revised downward we have to conclude that N has been enriched
in these stars relative to the \HII \ region values.

A better comparison of the stellar abundances with models of Galactic chemical evolution
require  better analyses
of the observational data, it is beyond the scope
of this paper to try to homogenize the available observational data.

\subsection{Comparison between models and observations}
  
The models were built to reproduce the O/H gradient and the present O/H
value of the ISM presented in Paper I. 
Models 1 and 2 predict for the time
the Sun was formed a value of 12 + log O/H =
8.66 for the ISM of the solar vicinity; in excellent agreement with 8.66 $\pm$ 0.05
the Asplund et al. (2005) value. Since the Sun was formed  the increase
in the O/H value of the ISM predicted by Models 1 and 2 
amounts to 0.13 dex, in excellent agreement with
the observations (see Table 5).

Of the 11 computed models only Models 1 and 2 are able
to reproduce: the C/H gradient, the present C/H value in the
solar vicinity, and the C/H solar value by Asplund et al. (2005). 
The other 9 models do not agree with the C observations.
The 12 + log(C/H)  predicted values by Models 1 and 2 for the solar vicinity at the 
time the Sun was formed
amount to 8.38 and 8.36, while those for log(C/O) amount to  $-0.28$ and $-0.30$ in very 
good agreement with the Asplund et al. values that amount to
8.39 $\pm$ 0.05 and $-0.27 \pm$ 0.05, respectively (see Table 5).
The predicted values for 12 + log(C/H) at present for the ISM amount to 8.67
and 8.62  for Models 1 and 2, in very good agreement with 8.67 $\pm$ 0.07, the observed value. 
The predicted 
values by  Models 1 and 2, for log(C/O) are $-0.12$ and $-0.17$,
while the present value derived in Paper I amounts to $-0.10$ $\pm$ 0.08.

The predicted C/O versus O/H
enrichment history for the solar vicinity of Models 1 and 2 is in
reasonable agreement with 
metal poor unevolved dwarfs, the Solar values, and the values of metal rich stars
of the solar vicinity (see Figure 3).
The behavior of Model 1 for the halo is similar to that by Akerman et al. (2004).
The increase of C/O with O/H is due to two factors:
i)  the bulk of metal poor LIMS eject large amounts of C in the
 $7.9 < $ 12+log(O/H) $ < 8.2$ range, and
ii)  metal rich massive stars eject more C than O for 12+log(O/H) $> 8.2$.

Akerman et al. (2004)  were the first to present
observational  evidence that log(C/O) $\sim -0.6$ for the very metal poorest stars 
with  12+log(O/H) $\sim 6$; and that for  stars in the $ 6< $ 12+log(O/H) $ < 7.6 $ range
the log(C/O) value diminished monotonically  to $-0.95$. 
The model by Akerman et al. is able to explain the decrease of C/O
from $-0.6$ dex to $-0.75$ dex because they have assumed
the yields by Chieffi \& Limongi (2002) for $Z=0$.
 Among Pop III yields, those by Chieffi \& Limongi  present the highest
C/O yields ratios. Then  the predicted C/O values decrease
because the C/O yields ratios by MM02  for $Z=10^{-5}$, $Z=0.001$, and $Z=0.004$
are lower than those for Pop III.
As mentioned in Akerman et al. (2004) the minimum C/O value given by the halo
objects has not been fitted by previous models nor by the models
presented in this paper.

There has been a discussion in the literature on the relative importance
of MS and LIMS in the C production by different authors.
While Henry, Edmunds, \& K\"oppen (2000)
 find that MS produce most of the C in the solar vicinity,
we find that MS and LIMS produce roughly the same fraction.
Our results  are different because:
a) Henry et al. adopted the C M92 yields for MS increased by factors in the 1.7 to 1.9
range, while we used the yields as presented  by MM02 and M92, and
b) Henry et al. adopted the Salpeter IMF that predicts a higher
number of MS than the KTG IMF adopted by us.

Other authors have discussed the contribution of MS and LIMS to the C enrichment,
but their results are not directly comparable to ours.
Carigi (2000) only gave a qualitative
description of the problem while in this paper we present precise percentages.
Carigi (2003) based on the M92 yields presented the instantaneously ejected C
as a function of time, including the net C produced by the stars
and the amount of initial C that was not converted into heavier atoms;
Akerman et al. (2004) based on the MM02 yields
presented only the synthesized and instantaneously ejected C as a function
of time; while we present only the cumulative synthesized C  up to a given time.

Gavil\'an et. al (2005) present new models of the C and O chemical
evolution of the solar vicinity and the Galactic disk, their models are similar
to our models 10 and 11 that predict a flat C/O vs $r$ behaviour.
As discussed in Sec 4.1 the flatness is mainly due to the adoption of
WW yields for MS. Our best models (1 and 2) do predict a C/O gradient
similar to the observed one.

Based on the observed N/H, N/O, and N/C abundances we find that:
Model 1 is in agreement with the \HII \ regions  and the metal poor stars,
Model 2 is in better agreement with the metal rich stars than Model 1, 
Model 2 does not fit the \HII \ regions and the metal-poor stars,
and the solar values
are intermediate between Models 1 and 2
(see Table 5 and Figure 5).

With the combined contribution of Type Ia and Type II SNe
 Models 1 and 2 predict Fe/H values of 7.46 dex and 7.49 dex
for the ISM in the solar vicinity at the time the Sun was formed, in excellent
agreement with the Asplund et al. (2005)
solar value.
The present Fe/H values predicted by Models 1 and 2 are 7.72 dex and 7.75 dex, respectively.

The C/Fe and O/Fe solar values are well fitted by Models 1 and 2 (see Figure
7). Alternatively the other stars have on average values about 0.3 dex
higher than those predicted by Models 1 and 2.
 Probably part of the
discrepancy is due to difference in the assumptions made 
in  the solar determination by Asplund et al. (2005) and those assumptions 
made in the determinations  for other stars. 

It is possible to lower the Fe production predicted by Models 1 and 2,
without affecting  the C/O versus O/H history,
either by the reducing the fraction of the stars that produce Type Ia SNe
or by adopting the Fe B yields by WW95 for Type II SNe.
In this case the models will produce a better fit to the C/Fe and O/Fe
versus Fe/H diagrams for the stars of the solar vicinity,
but the predicted solar C/Fe and O/Fe values would become higher
than the observed ones.
That is, either the Sun is C/Fe and O/Fe poor relative to the other stars,
or the C/Fe and O/Fe abundances determinations for the solar vicinity stars
present systematic effects relative to the solar determinations by Asplund et al. (2005).

Some of our results might apply to other galaxies. Nevertheless, we like to 
point out that the star formation history for other galaxies might be different to that of the Galaxy
and that tailor made models for representative galaxies are needed to see
if our results can be generalized to other objects.

\subsection{The use of a Salpeter like IMF}

The results presented previously depend on the IMF adopted.
Therefore to study the effect of a different IMF we decided
to run chemical evolution models with a modified IMF for $m > 1 \msun$.

For $m > 1 \msun$ the KTG IMF shows a very similar slope than that presented
by Scalo (1986, with $\alpha=-2.63$) and a steeper slope than that presented
by Salpeter (1955, with $\alpha=-2.35$).
We will adopted a ``Salpeter like" IMF defined as follows:
a slope $\alpha=-2.35$ for $m > 1 \msun$ only and the KTG slopes for $m < 1 \msun$.

Since oxygen is produced primarily by massive stars,
the computed oxygen abundances are very sensitive to
the behaviour of the IMF in the high mass end.
A model with a Salpeter like IMF and a mass upper limit ($m_{up}$) of 80 \msun \ predicts 
12 + log(C/H)$ = 8.86$  and 12 + log(O/H)$ = 8.95$  for the solar values 
while  for the present time ISM the predicted values are 9.13 and 9.08, respectively;
these values are considerably higher than observed.
Moreover the predicted gas mass (18 \msun $pc^{-2}$) 
is higher than the observed one (13 $\pm 3$ \msun $pc^{-2}$).
In order to improve the  match, we computed several models reducing
the upper limit masses of the IMF down to 40 \msun. The resulting values
for $m_{up}=40 $ \msun are 12 + log(C/H)$ = 8.54$ and 12 + log(O/H) $ = 8.76$ for solar values
and 8.81 and 8.94 for the present time ISM values,
these values  are still considerably higher than the observed ones.
Moreover the predicted gas mass (17 \msun $pc^{-2}$) is still higher than the observed one.

Similarly all models  based on a Salpeter like IMF with reasonable $m_{up}$ values 
($m_{up} > 40$ \msun)  fail also to
reproduce the behavior of the C/O vs O/H enrichment history
and the observed gradients simultaneously.

For $m_{up}=40 $ \msun \
the predicted and observed slopes of the C/H gradients are similar, but
 the predicted slope of the O/H gradient is steeper,
and the predicted slope of the C/O gradient is flatter than the observed ones.

The results of this section imply that the observational constraints are strong enough
to permit to discriminate among different IMFs.
In a future paper we plan to use different IMFs to study exhaustively
their effects on the chemical evolution of the solar vicinity and the Galactic disk.

\section{Conclusions}

We present a set of 11 models built to reproduce the observed O/H gradient
including the present solar vicinity O/H value of the ISM.
We tested these models with other observational constraints
and the main results are presented below.

Models 1 and 2 predict C/Fe and O/Fe versus Fe/H values from 0 to 0.4 dex
smaller than observed. On the other hand, they produce
a very good fit to the C/Fe, O/Fe solar values.
For the N/Fe versus Fe/H enrichment history Models 1 and 2 differ considerably
and the stellar data fall in between. Model 2 is closer to the
solar value than Model 1.

There is an observational discrepancy between the  N/C  and N/O
values derived from \HII \ regions and those derived from metal-rich stars
of the solar vicinity that 
has to be sorted out before definitive conclusions of the 
Galactic N enrichment history are  reached. 
While Model 1 produces an excellent agreement with the N/C and N/O values
derived from observations of \HII \ regions and with the N/O values derived
from metal poor stars, Model 1 is not able to produce a good fit for the N/O 
values derived
from metal rich stars. Alternatively Model 2 produces a fair fit to the N/C and N/O
values derived from metal rich stars, but fails to explain the \HII \ region
observations as well as those of metal poor stars.
Consequently the N enrichment problem has to be studied further.
A more precise discussion of the N enrichment history 
requires a solution of the observational discrepancy
in the N/C and N/O values 
between the \HII \ regions and the metal rich stars,
as well as a revision of the N yields available.

Models 1 and 2 predict an enrichment in the O/H ratio of 0.13 dex since
the Sun was formed. By adding this value to the Asplund et al. (2005)
12 + log O/H = 8.66
solar value,  we predict for the ISM in the solar vicinity a value of
12 + log O/H = 8.79 in excellent agreement with the value derived from the
recombination line observations, after correcting by the dust presence,
that amounts to 12 + log O/H = 8.77 $\pm$ 0.05.

Models 1 and 2 predict an increase in the C/H ratio of 0.29 dex and 0.26 dex since
the Sun was formed. By adding these values to the Asplund et al. (2005)
12 + log C/H = 8.39
solar value,  we predict for the ISM in the solar vicinity  values of
12 + log C/H = 8.68 and 8.65 in excellent agreement with the value derived from the
recombination line observations, after correcting by the dust presence,
that amounts to 12 + log C/H = 8.67 $\pm$ 0.07.

From the values presented in Table 5 and Figure 3, it is clear that 
Model 1 produces a better fit
to the \HII \ region restrictions, to the solar values, and to
the C/O versus O/H enrichment history than Model 2.

In this paper we present a solution to the C enrichment
history of the Galaxy based on the yields
and observations available.
The solution is based on the adoption of C yields that increase with metallicity
due to stellar winds in MS and decrease with metallicity 
due to stellar winds in LIMS.
These yields fit the behavior of the C/O ratio in the $6 < r({\rm kpc}) < 11$ 
range, the range for which we have C/H and O/H values from \HII \ regions
based on recombination  lines. The adopted yields also produce a reasonable
fit to the C/O history of the solar vicinity.

We also find that about half of the C in the ISM of the solar vicinity at the present time
has been produced by MS and half by LIMS. Also, at the present time,
for  a Galactocentric distance of 6 kpc about 53 \% of the C has been produced by MS
and  45 \% by LIMS, while for 11 kpc the opposite is true,
about 42 \% of the C has been produced by MS and 56 \% by LIMS.

It is clear that a more powerful treatment of convection,
a better value of the  $^{12}$C($\alpha$,$\gamma$)$^{16}$O rate, 
and a more realistic mass loss rate scheme
will produce a better solution to the C enrichment problem.
Moreover to produce a more stringent test for the C yields 
it is necessary to obtain observations of the C/O ratio   for $r<  6$ kpc
and $ r> 11$ kpc and to include 
a model of the bulge formation and its effect on the C/O values for the
inner regions of the Galaxy.

We would like to thank Chris Akerman for providing us the observational
errors of the C, O and Fe  stellar values, as well as John
Bahcall for relevant correspondence. We would like to thank
also Dick Henry, the referee of this paper, for some excellent suggestions.
We also acknowledge John Scalo for suggesting us the use of another IMF.
LC and MP  received partial support from CONACyT (grant 36904-E) and
DGAPA UNAM (grant IN114601), respectively. 
CE and JGR received partial support from the
Spanish Ministerio de Ciencia y Tecnolog\'\i a (MCyT) under projects AYA2001-0436
and AYA2004-07466.

\clearpage

\begin{deluxetable}{cccc}
\tablecaption{N abundances in Galactic \HII \ regions \tablenotemark{a}
\label{ndata}}
\tablewidth{0pt}
\tablehead{
\colhead{\HII \ region } &
\colhead{$r$ (kpc)} &
\colhead{$t^2$} &
\colhead{12 + log(N/H)}
}
\startdata
M 16      & 6.34  & 0.036 $\pm$ 0.006 & 8.07 $\pm$ 0.12\\
M 8       & 6.41  & 0.037 $\pm$ 0.004 & 7.94 $\pm$ 0.06\\
M 17      & 6.75  & 0.033 $\pm$ 0.005 & 7.87 $\pm$ 0.13\\
M 20      & 7.19  & 0.036 $\pm$ 0.013 & 7.89 $\pm$ 0.09\\
NGC 3576  & 7.46  & 0.038 $\pm$ 0.009 & 7.87 $\pm$ 0.09\\
Orion neb.& 8.40  & 0.022 $\pm$ 0.002 & 7.73 $\pm$ 0.15\\
NGC 3603  & 8.65  & 0.040 $\pm$ 0.008 & 7.89 $\pm$ 0.15\\
S 311     &10.43  & 0.038 $\pm$ 0.007 & 7.61 $\pm$ 0.07\\
\enddata
\tablenotetext{a}{
Gaseous abundances. Galactocentric distances from Paper I.
}
\end{deluxetable}

\begin{deluxetable}{cccccc}
\tablecaption{Present-day radial gradients
\label{grad}}
\tablewidth{0pt}
\tablehead{
&
\multicolumn{3}{c}{Assumed Yields} &
\multicolumn{2}{c}{C/O} \\
&
\colhead{MS}  &
\colhead{MS}  &
\colhead{LIMS}  &
\colhead{value (dex)} &
\colhead{slope (dex kpc$^{-1}$)} \\
\colhead{Model} &
\colhead{$0 < Z < 0.02$} &
\colhead{$Z \ge 0.02$ } &
&
\colhead{$ r=r_\odot=8$ kpc} &
\colhead{$6 < r/{\rm kpc} < 11$} 
}

\startdata
1 & MM02 & M92 & MBCP      & $- 0.122$ & $- 0.057$ \\
2 & MM02 & M92 & vdHG.var  & $- 0.172$ & $- 0.053$ \\
3 & MM02 & M92 & vdHG.const  & $- 0.249$ & $- 0.066$ \\
4 & MM02 & M92 & MM02      & $- 0.350$ & $- 0.068$ \\
& & & & & \\
5 & MM02 & MM02 & MBCP     & $- 0.279$ & $- 0.025$ \\
6 & MM02 & MM02 & vdHG.var & $- 0.355$ & $- 0.012$ \\
7 & MM02 & MM02 & vdHG.const & $- 0.410$ & $- 0.029$ \\
8 & MM02 & MM02 & MM02     & $- 0.542$ & $- 0.024$ \\
& & & & & \\
9 & PCB98 & PCB98  & MBCP   & $- 0.142$ & $- 0.004$\\
10 & WW95 & WW95 & vdHG.var & $- 0.163$ & $- 0.014$\\
11 & WW95 & WW95 & vdHG.const & $- 0.463$ & $- 0.005$ \\
& & & & & \\
Obs \tablenotemark{a} & & & & $- 0.102 \pm 0.080$ & $- 0.058 \pm 0.020$   \\
\enddata
\tablenotetext{a}{ Paper I values corrected by dust, see section 2}

\end{deluxetable}

\begin{deluxetable}{cccc}
\tablecaption{Carbon budget for the solar vicinity \tablenotemark{a}
\label{cprodsv}}
\tablewidth{0pt}
\tablehead{
&
\multicolumn{3}{c}{Contribution (per cent)} \\
\colhead{Model} &
\colhead{MS} &
\colhead{LIMS} &
\colhead{SNIa}
}
\startdata
1 & 48.2 &  49.8 & 2.0 \\
2 & 50.2 &  47.2 & 2.6 \\
3 & 59.5 &  38.0 & 2.5 \\
4 & 75.3 &  21.8 & 2.9 \\
&  & & \\
5 & 40.2 &  57.5 & 2.3 \\
6 & 40.8 &  56.4 & 2.8 \\
7 & 52.7 &  44.4 & 2.9 \\
8 & 70.0 &  26.1 & 3.9 \\
&  & & \\
9 & 51.8 &  47.0 & 1.2 \\
10  & 36.2 &  62.3 & 1.5 \\
11 & 66.7 &  32.1 & 1.9 \\
\enddata
\tablenotetext{a}{
Percentage of C in the ISM
produced by different types of stars 
over a period of 13 Gyr.
}
\end{deluxetable}

\begin{deluxetable}{cccc}
\tablecaption{Carbon budget for different Galactocentric distances \tablenotemark{a}
\label{cprodr}}
\tablewidth{0pt}
\tablehead{
&
\multicolumn{3}{c}{Contribution (per cent)} \\
\colhead{$r$ (kpc) } &
\colhead{MS} &
\colhead{LIMS}&
\colhead{SNIa}
}
\startdata
&
\multicolumn{3}{c}{Model 1} \\
 4 & 57.0 & 40.8 & 2.2 \\
 6 & 53.4 &  44.6 & 2.0 \\
8  & 48.2 &  49.8 & 2.0 \\
 11 & 42.4 &  55.5 & 2.1 \\
 16 & 40.8 & 57.2 & 2.1 \\
&  & & \\
&
\multicolumn{3}{c}{Model 2} \\
 4 & 63.1 & 34.0 & 2.9 \\
 6  &  58.0 & 39.3 & 2.7 \\
8 & 50.2 &  47.2 & 2.6 \\
 11 &  38.6 & 59.0 & 2.4 \\
 16 & 30.3 & 67.6 & 2.1 \\
\enddata
\tablenotetext{a}{
Percentage of C in the ISM
produced by different types of stars
over a period of 13 Gyr.
}
\end{deluxetable}

\begin{deluxetable}{cccccc}
\tablecaption{ISM abundance values \tablenotemark{a}
\label{ismabund}}
\tablewidth{0pt}
\tablehead{
\colhead{ } &
\colhead{O/H} &
\colhead{C/H}&
\colhead{N/H} &
\colhead{Fe/H} &
\colhead{C/O}
}
\startdata
&
\multicolumn{5}{c}{At the time the Sun was formed ($t = 8.43$ Gyr)} \\
 Model 1  & 8.66 & 8.38 & 7.56 & 7.46 & $- 0.28$  \\
 Model 2  & 8.66 & 8.36 & 7.89 & 7.49 & $- 0.30$  \\
 Solar \tablenotemark{b}  & 8.66 $\pm$ 0.05 & 8.39 $\pm$ 0.05 & 7.78 $\pm$ 0.05 
& 7.45 $\pm$ 0.05 & $- 0.27 \pm$ 0.07  \\
& & &  & & \\
&
\multicolumn{5}{c}{At the present time ($t = 13.0$ Gyr)} \\
 Model 1  & 8.79 & 8.67 & 7.84 & 7.72 & $- 0.12$  \\
 Model 2  & 8.79 & 8.62 & 8.13 & 7.75 & $- 0.17$  \\
 \HII \ Regions & 8.77 $\pm$ 0.05 \tablenotemark{c} 
& 8.67 $\pm$ 0.07 \tablenotemark{c} & 7.84 $\pm$ 0.10 \tablenotemark{d} & $--$ &
 $- 0.10 \pm$ 0.08 \tablenotemark{c} \\
\enddata
\tablenotetext{a}{ Given in 12 + log (X/H).}
\tablenotetext{b}{ Asplund et al. (2005).}
\tablenotetext{c}{ Paper I values corrected by dust. }
\tablenotetext{d}{ This paper. }
\end{deluxetable}

\begin{deluxetable}{cccc}
\tablecaption{Nitrogen budget for different Galactocentric distances \tablenotemark{a}
\label{nprod}}
\tablewidth{0pt}
\tablehead{
&
\multicolumn{3}{c}{Contribution (per cent)} \\
\colhead{$r$ (kpc) } &
\colhead{MS} &
\colhead{LIMS}&
\colhead{SNIa}
}
\startdata
&
\multicolumn{3}{c}{Model 1} \\
4  & 44.5 & 53.5 & 2.0  \\
 6  & 40.2 & 58.0 & 1.8  \\
 8  & 33.5 & 64.7 & 1.8  \\
 11 & 23.6 & 74.5 & 1.8  \\
 16 & 12.0 & 86.0 & 2.9  \\
&  & & \\
&
\multicolumn{3}{c}{Model 2} \\
 4  & 26.6 & 68.8 & 4.7  \\
 6  & 23.7 & 72.0 & 4.3  \\
 8  & 19.2 & 76.7 & 4.1  \\
 11 & 12.7 & 83.6 & 3.7  \\
 16 & 6.3  & 90.1 & 3.2  \\
\enddata
\tablenotetext{a}{
Percentage of N in the ISM
produced by different types of stars
over a period of 13 Gyr.
}
\end{deluxetable}

\newpage

\begin{figure}[f1]
\plotone{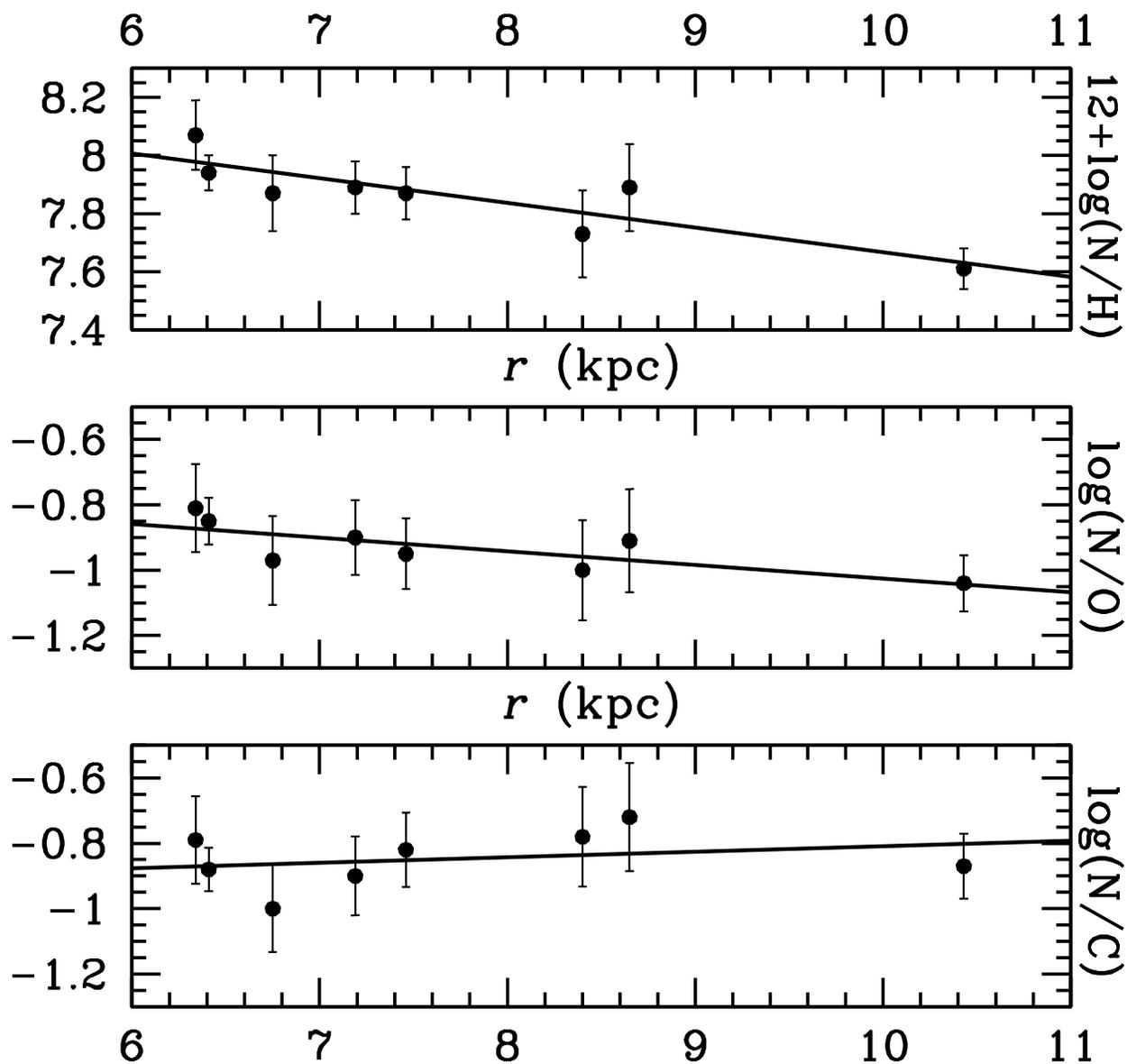}
\caption{
N/H, N/C, and N/O radial abundance  gradients of the Galactic disk derived
from \HII \ regions. The N abundances have been determined from collisionally
excited lines while those of H, C and O have been obtained from recombination
lines (see Table 1 and Paper I). The lines indicate the least-squares linear 
fits to the data.
}
\end{figure}

\begin{figure}[f2]
\plotone{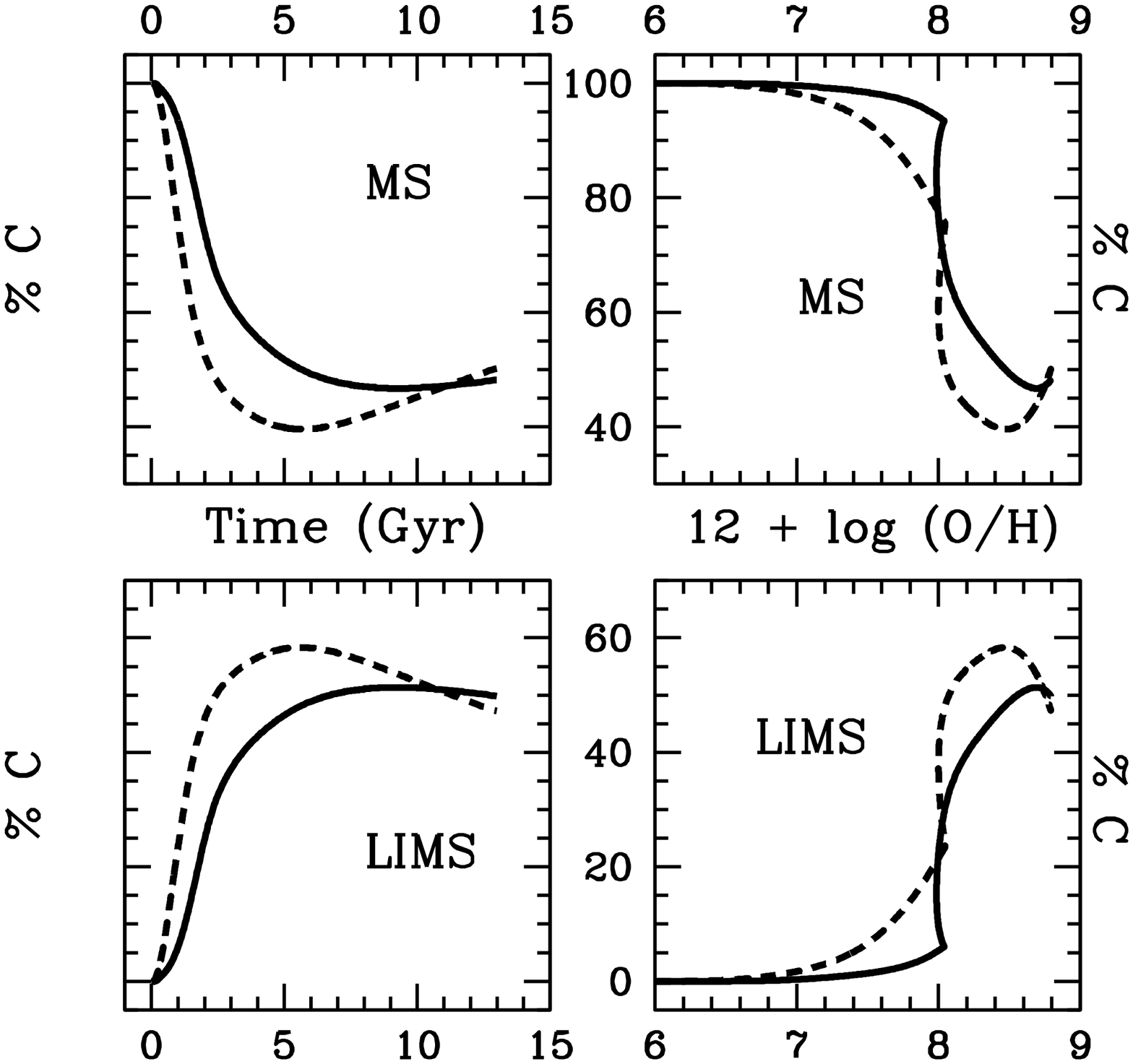}
\caption{
Cumulative percentage of C as a function of time and 12 + log(O/H),
due to massive stars (MS), and low and intermediate-mass stars (LIMS)
at the solar vicinity.
Solid lines represent Model 1
and broken lines represent Model 2.
}
\end{figure}

\begin{figure}[f3]
\plotone{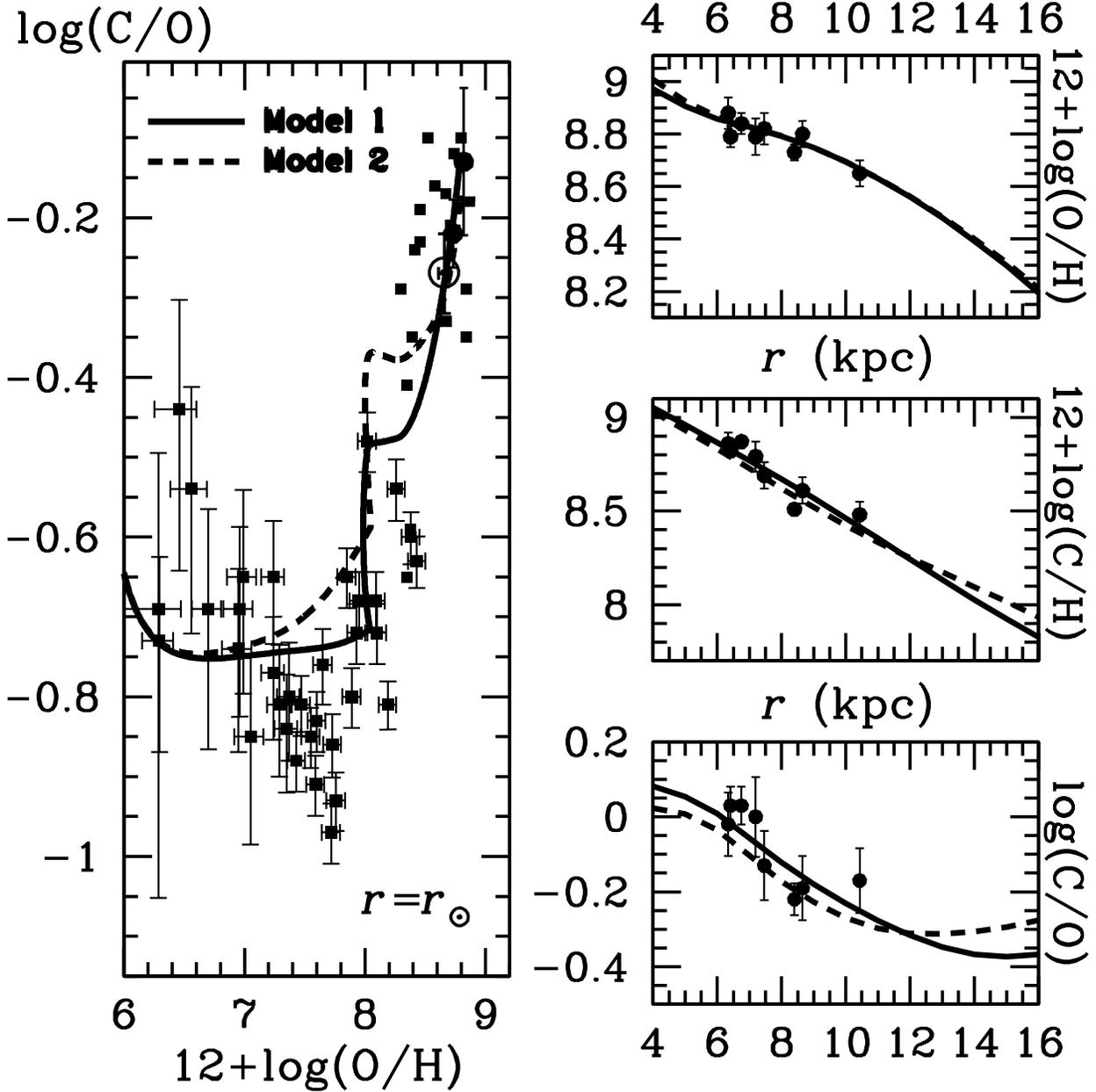}
\caption{
Predictions from models considering: a) for massive stars, 
yields by Chieffi \& Limongi (2002) for $Z = 0.0$,
yields by  Meynet \& Maeder (2002) for $10^{-5} < Z < 0.004$, Maeder (1992) for $Z = 0.02$;
and, b) for low and intermediate mass stars,
yields by Marigo et al. (1996, 1998) and Portinari et al. (1998) (Model 1), or
van den Hoek \& Groenewegen (1997)  with $\eta$ variable (Model 2).
The left panel shows the C/O evolution in the ISM of the solar vicinity with O/H.
The right panels show the present-day ISM abundance ratios as a function 
of Galactocentric distance.
{\it Filled circles:} \HII \ regions,  gas plus dust values; the gaseous values from Paper I
have been corrected by the dust fraction (see section 2).
{\it Filled squares:} dwarf stars  from Akerman et al. (2004).
{$\odot$:} Solar values from Asplund et al. (2005).
}
\end{figure}

\begin{figure}[f4]
\plotone{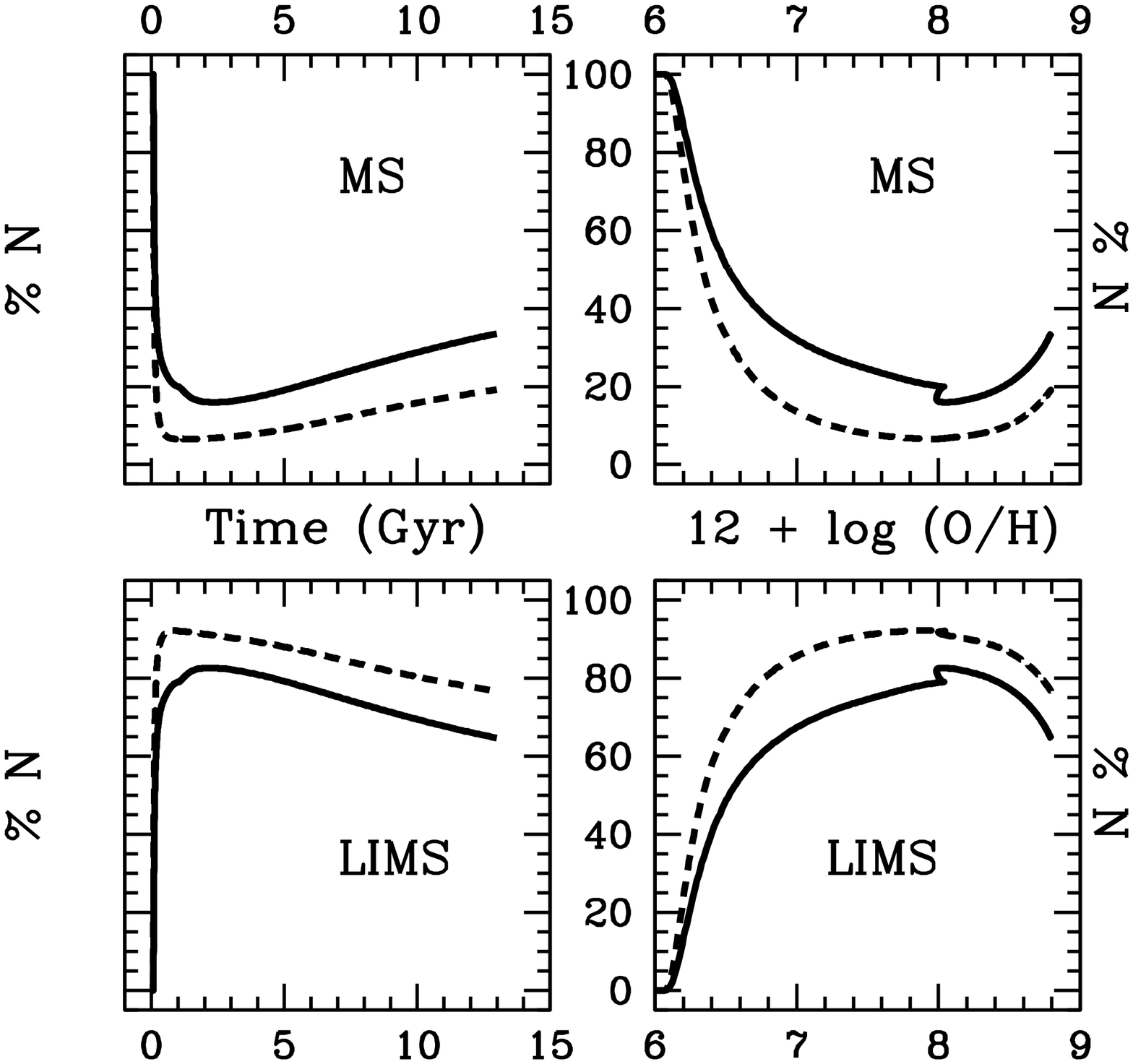}
\caption{
Cumulative percentage of N as a function of time and 12 + log(O/H),
due to massive stars (MS), and low and intermediate-mass stars (LIMS)
at the solar vicinity.
Solid lines represent Model 1
and broken lines represent Model 2.
}
\end{figure}

\begin{figure}[f5]
\plotone{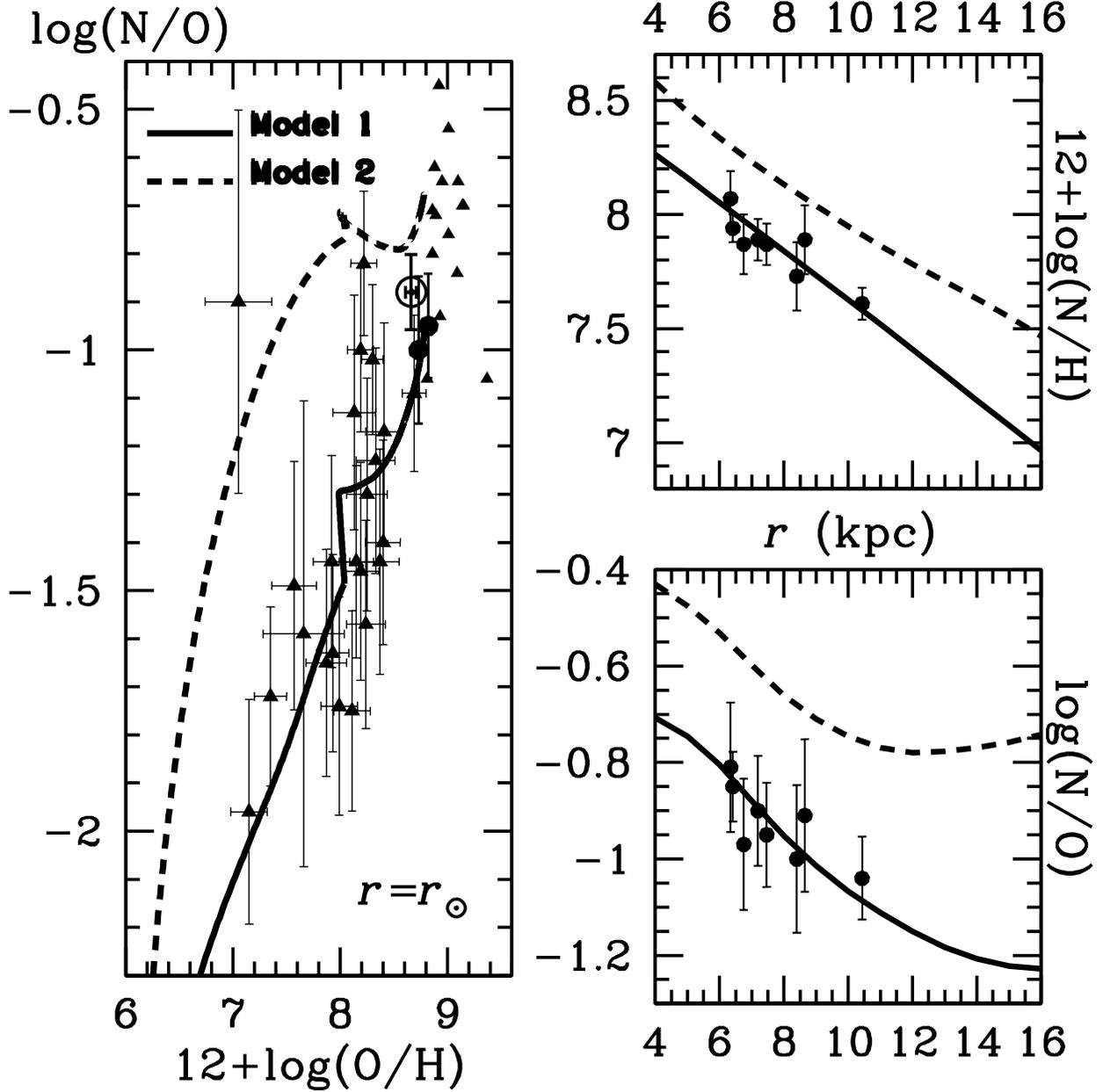}
\caption{
N/H and N/O gradients in the Galactic disk and 
N/O versus O/H enrichment history in the solar vicinity.
The left panel shows the N/O evolution in the ISM of the solar 
vicinity with O/H.
The adopted N yields for $Z = 0.02$ are those by 
Meynet \& Maeder (2002), all the other yields are as in Fig. 3
(see section 3). 
The right panels show the present-day ISM abundance ratios as a
 function of Galactocentric distance.
{\it Filled circles:} \HII \ region values presented in Fig. 1,
{\it Filled triangles:} unevolved main sequence stars 
from Israelian  et al. (2004),
{$\odot$:} solar values from Asplund et al. (2005).
}

\end{figure}

\begin{figure}[f6]
\plotone{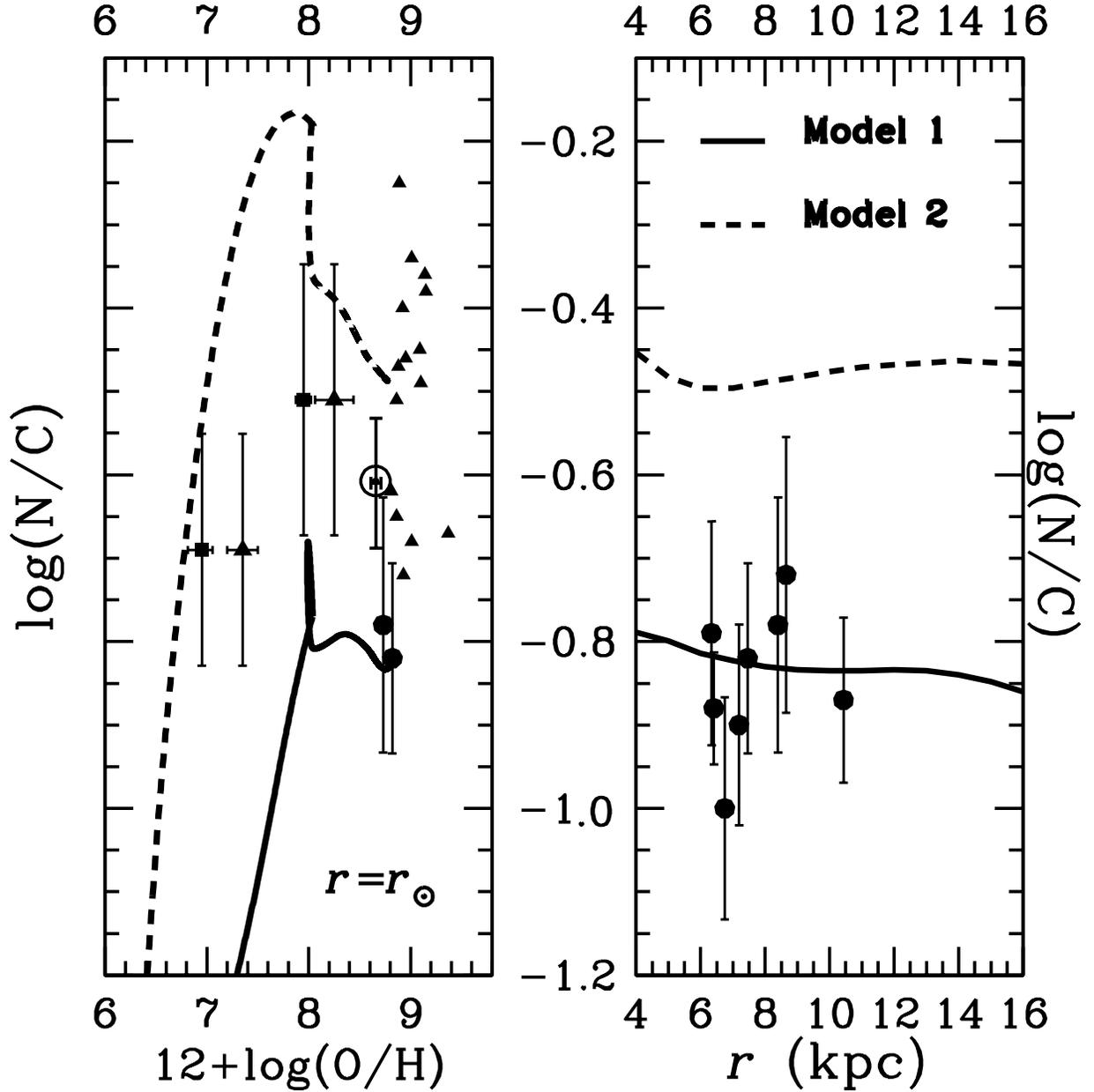}
\caption{
N/C  gradients in the Galactic disk and the
N/C versus O/H enrichment history in the solar vicinity.
The left panel shows the N/C evolution in the ISM of the solar
vicinity with O/H.
The right panel shows the present-day ISM abundance ratios as a
function of Galactocentric distance.
{\it Filled circles:} \HII \ region values presented in Fig. 1,
{\it filled triangles:} unevolved stars from
Santos et al. (2000), Takeda et al. (2001),
Gonz\'alez et al. (2001), Sadakane et al. (2002), and
Israelian  et al. (2004).
{$\odot$:} solar values from Asplund et al. (2005).
}

\end{figure}

\begin{figure}[f7]
\plotone{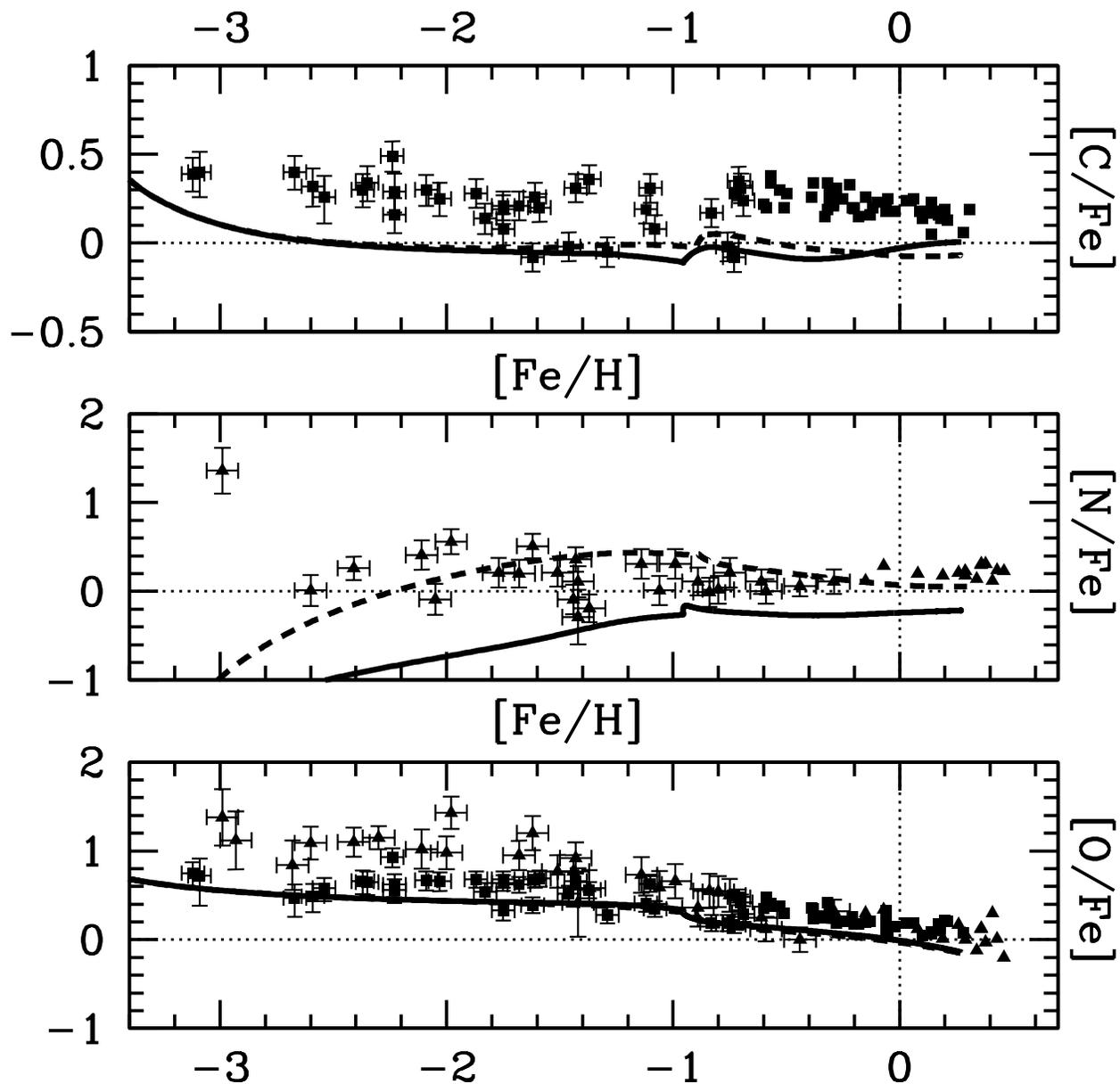}
\caption{
[C/Fe], [N/Fe], and [O/Fe] vs [Fe/H]  in the solar vicinity.
{\it Filled triangles:} Unevolved stars from
Israelian  et al. (2004),
{\it filled squares:} dwarf stars from Akerman et al. (2004).
Solid lines represent Model 1
and broken lines represent Model 2.
}
\end{figure}

\end{document}